\documentclass[aps,prd,groupedaddress]{revtex4}
\usepackage{graphicx}
\usepackage{epsfig}
\usepackage{amsfonts}
\usepackage{amssymb}
\usepackage{epsf}
\usepackage{xcolor}
\newcommand{\insertplot}[5]{\begin{figure}
 \hfill\hbox to 0.05in{\vbox to #5in{\vfill
 \inputplot{#1}{#4}{#5}}\hfill}
 \hfill\vspace{-.1in}
 \caption{#2}\label{#3}
 \end{figure}}
\newcommand{\inputplot}[3]{
 \special{ps: plotfile #1}
}

\newcounter{fig}

\def\beq{\begin{equation}}
\def\eeq{\end{equation}}
\def\bea{\begin{eqnarray}}
\def\eea{\end{eqnarray}}

\begin{document}
\title{Novel Einstein-Scalar-Gauss-Bonnet Wormholes without Exotic Matter}
\author{Georgios Antoniou}
\email[]{anton296@umn.edu}
\affiliation{School of Physics and Astronomy, University of Minnesota,
             Minneapolis, MN 55455, USA}
\author{Athanasios Bakopoulos}
\email[]{abakop@cc.uoi.gr}
\author{Panagiota Kanti}
\email[]{pkanti@cc.uoi.gr}
\affiliation{Division of Theoretical Physics, Department of Physics,
             University of Ioannina, Ioannina GR-45110, Greece}
\author{Burkhard Kleihaus}
\email[]{b.kleihaus@uni-oldenburg.de}
\author{Jutta Kunz}
\email[]{jutta.kunz@uni-oldenburg.de}
\affiliation{Institut f\"ur Physik, Universit\"at Oldenburg,D-26111 Oldenburg, Germany}

\date{\today}
\begin{abstract}
Novel wormholes are obtained in Einstein-scalar-Gauss-Bonnet theory
for several coupling functions. The wormholes may feature a single-throat or
a double-throat geometry and do not demand any exotic matter.
The scalar field may asymptotically vanish or be finite, and it may possess radial excitations.
The domain of existence is fully mapped out for various forms of the
coupling function.
\end{abstract}

\maketitle


\section{Introduction}

In the quest for the fundamental theory of gravity, 
Einstein-scalar-Gauss-Bonnet
(EsGB) theories represent interesting alternative theories of gravity
(see e.g.~\cite{Berti:2015itd,Sotiriou:2015lxa}).
They belong to the class of {\sl quadratic gravitational theories} 
that contain higher-curvature gravitational terms. 
These terms are treated as small deformations
that nevertheless complete Einstein's General Relativity 
and may modify its predictions at regimes of strong gravity. 
In the EsGB theory, the Einstein-Hilbert action is supplemented 
by a scalar field, non-minimally coupled to the
quadratic Gauss-Bonnet (GB) term. 
The resulting field equations are of second order,
avoiding Ostrogradski instability and ghosts 
\cite{Horndeski:1974wa,Charmousis:2011bf,Kobayashi:2011nu}.
In addition, this quadratic theory has so far survived the constraints 
set by the detection of gravitational waves emitted during the
binary mergers, when the coupling function allows to set the
scalar field to zero in the cosmological context,
and thus lead to the same solutions as the standard cosmological
$\Lambda$CDM model \cite{Sakstein:2017xjx}.
The study of the types of solutions that this theory admits 
is therefore of paramount importance.
 
Motivated by string theory with the dilaton as the scalar field,
the Einstein-dilaton-Gauss-Bonnet (EdGB) theory features an exponential
coupling between the scalar field and the GB term
\cite{Zwiebach:1985uq,Gross:1986mw,Metsaev:1987zx}.
Black-hole solutions arising in the context of the EdGB theory differ
from the Schwarzschild or Kerr black holes since they possess a non-trivial
dilaton field and thus carry dilaton hair 
\cite{Kanti:1995vq,Torii:1996yi,Guo:2008hf,Pani:2009wy,Pani:2011gy,Kleihaus:2011tg,Ayzenberg:2013wua,Ayzenberg:2014aka,Maselli:2015tta,Kleihaus:2014lba,Kleihaus:2015aje,Blazquez-Salcedo:2016enn}.
The extended family of EsGB theories, where different coupling functions of
the scalar field to the GB term may be employed, has attracted recently
considerable attention 
\cite{Sotiriou:2013qea,Sotiriou:2014pfa,Antoniou:2017acq,Doneva:2017bvd,Silva:2017uqg,Antoniou:2017hxj,Witek:2018dmd,Minamitsuji:2018xde,Silva:2018qhn,Bakopoulos:2018nui,Brihaye:2018grv,Macedo:2019sem,Doneva:2019vuh,Myung:2019wvb,Cunha:2019dwb,Brihaye:2019dck}.
In these theories, new black holes arise through {\sl spontaneous} 
or {\sl induced scalarization} depending on whether the scalar field 
acquires a zero or non-zero, respectively, value at infinity.
The stability of scalarized black holes of EsGB theories 
has been addressed in detail by
analyzing their radial perturbations and revealing a distinct dependence on
the coupling function \cite{Blazquez-Salcedo:2018jnn}. 

A particularly interesting property emerging in the EdGB solutions is the
presence of regions with negative {\sl effective} energy density  -- this
is due to the presence of the higher-curvature GB term and is therefore
of purely gravitational nature \cite{Kanti:1995vq,Kanti:2011jz}.
Consequently, the EdGB theory allows for Lorentzian, traversable wormhole
solutions without the need for exotic matter \cite{Kanti:2011jz,Kanti:2011yv}.
It is tempting to conjecture that the more general EsGB theories should also
allow for traversable wormhole solutions. 
Indeed, traversable wormholes require violation of the energy conditions
\cite{Morris:1988cz,Visser:1995cc}.
But whereas in General Relativity this violation is typically 
achieved by a phantom field
\cite{Ellis:1973yv,Ellis:1979bh,Bronnikov:1973fh,Kodama:1978dw,Kleihaus:2014dla,Chew:2016epf},
in EdGB theories it is the effective stress-energy tensor 
that allows for this violation \cite{Kanti:2011jz,Kanti:2011yv}.

Thus, in the context of this work, we consider
a general class of EsGB theories with an arbitrary coupling function for the
scalar field. We first readdress the case of the exponential coupling function,
and show that the EdGB theory is even richer than previously thought, since
it features also wormhole solutions with a double throat and an equator in
between. Then, we consider alternative forms of the scalar coupling function,
and demonstrate that the EsGB theories 
always allow for traversable wormhole solutions,
featuring both single and double throats. The scalar field may vanish or be
finite at infinity, and it may have nodes. 
We also map the domain of existence (DOE)
of these wormholes in various exemplifications, evaluate their global charges
and throat areas and demonstrate that the throat remains open without the
need for any exotic matter.

In section II we briefly recall EsGB theory and discuss the throat geometry
for single and double throat configurations. We here present the asymptotic
expansions near the throat/equator and in the two asymptotically flat regions,
and we also derive the formulae necessary to study the violation 
of the energy conditions.
We present our numerical solutions in section III, and discuss some of
their properties, their domains of existence and the energy conditions.
In order to impose symmetry of the solutions under reflection of the
spatial radial coordinate, $\eta \rightarrow -\eta$, we study 
in section IV the junction conditions at the throat/equator, 
and show that we can solve these by including a shell of ordinary matter,
only. In section V we present embedding diagrams of the wormhole solutions,
and we conclude in section VI. For completeness, we show the lengthy set of
field equations in appendix A.

\section{The Einstein-scalar-Gauss-Bonnet Theory}

We consider the following effective action describing a quadratic scalar-tensor theory
%
\begin{eqnarray}  
S=\frac{1}{16 \pi}\int d^4x \sqrt{-g} \left[R - \frac{1}{2}\,
 \partial_\mu \phi \,\partial^\mu \phi
 + F(\phi) R^2_{\rm GB}   \right].
\label{action}
\end{eqnarray} 
The theory contains the Ricci scalar $R$, a scalar field $\phi$ and the quadratic
gravitational Gauss-Bonnet term defined as
\beq
R^2_{\rm GB} = R_{\mu\nu\rho\sigma} R^{\mu\nu\rho\sigma}
- 4 R_{\mu\nu} R^{\mu\nu} + R^2 \label{GB}
\eeq
in terms of the Riemann tensor $R_{\mu\nu\rho\sigma}$, the Ricci tensor $R_{\mu\nu}$ and the Ricci scalar $R$.
The GB term, a topological invariant in four dimensions, is coupled to the scalar
field through a coupling function $F(\phi)$. The form of the latter will be left arbitrary,
therefore, our analysis will apply to a whole class of Einstein-scalar-GB theories
described by the action (\ref{action}). 

The Einstein and scalar field equations are obtained by variation of the action with
respect to the metric, respectively the scalar field, and have the form
\begin{equation}
G_{\mu\nu} =  T_{\mu\nu} \ , \ \ \ 
\nabla^2 \phi + \dot{F}(\phi)R^2_{\rm GB}  =  0 \ , 
\label{einfldeq}
\end{equation}
where the stress-energy tensor of the theory is given by the following expression
\begin{equation}
T_{\mu\nu} =
-\frac{1}{4}g_{\mu\nu}\partial_\rho \phi \partial^\rho \phi 
+\frac{1}{2} \partial_\mu \phi \partial_\nu \phi
-\frac{1}{2}\left(g_{\rho\mu}g_{\lambda\nu}+g_{\lambda\mu}g_{\rho\nu}\right)
\eta^{\kappa\lambda\alpha\beta}\tilde{R}^{\rho\gamma}_{\alpha\beta}\nabla_\gamma \partial_\kappa F(\phi).
\label{tmunu}
\end{equation}
Note that $T_{\mu\nu}$ receives contributions from the kinetic term of the scalar field but
also from the GB term, with the latter being non-trivial for a non-constant coupling function
$F(\phi)$ as expected. The dot above $F(\phi)$ in the scalar-field equation denotes the
derivative with respect to the scalar field, and we have used the definitions 
$\tilde{R}^{\rho\gamma}_{\alpha\beta}=\eta^{\rho\gamma\sigma\tau}
R_{\sigma\tau\alpha\beta}$ and $\eta^{\rho\gamma\sigma\tau}= 
\epsilon^{\rho\gamma\sigma\tau}/\sqrt{-g}$.

In this work, we consider only static, spherically-symmetric solutions of the field equations. 
To this end, we employ the following line-element
\begin{equation}
ds^2 
= -e^{f_0(\eta)}dt^2+e^{f_1(\eta)}\left\{d\eta^2
+\left(\eta^2+\eta_0^2\right)\left(d\theta^2+\sin^2\theta d\varphi^2 \right)\right\}\,,
\label{metric} 
\end{equation}
The substitution of the above metric in the Einstein and scalar field equations,
given in Eqs.~(\ref{einfldeq}), leads to three second-order and one 
first-order ordinary differential equations (ODEs) that are displayed in Appendix
\ref{App-equations}. Due to the Bianchi identity, only three of these four equations
are independent. In our analysis, we choose to solve the three second-order equations
while the first-order one will serve as a constraint on the unknown quantities
$f_0$, $f_1$ and $\phi$. 

\subsection{Single and Double-Throat Geometry}

In our previous analyses \cite{Kanti:2011jz, Kanti:2011yv}, we determined traversable wormhole solutions
by employing the following line-element
\begin{equation}
ds^2 
= -e^{f_0(l)}dt^2+p(l)\,dl^2
+\left(l^2+r_0^2\right)\left(d\theta^2+\sin^2\theta d\varphi^2 \right).
\label{metricSold} 
\end{equation}
The wormhole geometry is characterised by the circumferential radius $R_c$ defined as
\beq
R_c = \frac{1}{2\pi}\,\int_0^{2 \pi} \sqrt{g_{\varphi \varphi}}|_{\theta=\pi/2}\, d\varphi\,.
\label{circum-radius}
\eeq
A minimum of $R_c$ corresponds to a wormhole throat, whereas a local maximum corresponds to
an equator. For the line-element (\ref{metricSold}), the circumferential radius is
$R_c(l) = \sqrt{l^2+r_0^2}$.  This expression clearly possesses a minimum at $l=0$,
corresponding to a throat of radius $r_0$, and it does not allow for a local
maximum, i.e. an equator. Consequently, wormholes with only a throat and no equator
were presented in \cite{Kanti:2011jz,Kanti:2011yv}.

For the set of coordinates defined in Eq. (5), the circumferential radius
is expressed as $R_c(\eta) = e^{f_1/2}\sqrt{\eta^2+\eta_0^2}$. As we will see, this expression
allows for the existence of one or two local minima (i.e. throats) and of a local maximum
(i.e. equator). We introduce the distance variable in a coordinate-independent way as
\beq
\xi = \int_0^\eta \sqrt{g_{\eta \eta}}\,d\tilde \eta = \int_0^\eta e^{f_1(\tilde\eta)/2}\,d\tilde\eta \,.
\eeq
The conditions for a throat, respectively equator, at $\eta=0$ then read
\beq
\frac{dR_c}{d\xi}\Biggr|_{\eta=0}=0\,, \qquad \frac{d^2R_c}{d\xi^2}\Biggr|_{\eta = 0} \gtrless 0\,,
\label{cond-1}
\eeq
where the greater sign ($>$) refers to a throat and the smaller sign ($<$) to an equator. Using
the metric (5), these conditions yield
\beq
f_1'(0)=0\,, \qquad \eta_0^2 \,f_1''(0) + 2 \gtrless 0 \,. \label{cond-2}
\eeq
where the prime denotes derivative with respect to $\eta$. In the degenerate case, when the
throat and equator coincide, the inequalities in Eqs. (\ref{cond-1}) and (\ref{cond-2})
become equalities. If a throat/equator is located at  $\eta=0$, then its area is given by 
$A_{t,e} = 4\pi R^2_c(0)=4\pi \eta_0^2 e^{f_1(0)}$, while for a double-throat wormhole
with the throat located at $\eta_{\rm t}$, 
$A_{t} = 4\pi R^2_c(\eta_{\rm t})=4\pi (\eta_{\rm t}^2 + \eta_0^2 )e^{f_1(\eta_{\rm t})}$.


\subsection{Asymptotic Expansions}
\subsubsection{Expansion near the throat/equator}

A traversable wormhole solution is characterised by the absence of horizons or singularities.
In order to ensure that this is the case for our solutions, we consider the following
regular expansions for the metric functions and scalar field, near the throat/equator at
$\eta=0$,
\bea
    e^{f_0}&=&a_0\,(1+a_1 \eta +a_2 \eta^2+...)\,,\label{appf0}\\[1mm]
    e^{f_1}&=&b_0\,(1+b_1 \eta +b_2 \eta^2+...)\,,\label{appf1}\\[1mm]
    \phi&=&\phi_0+\phi_1 \eta +\phi_2 \eta^2+...\,.\label{appff}
\eea
The Lorentzian signature of spacetime demands that both parameters $a_0$ and $b_0$ must be positive;
in addition, they should be finite and non-vanishing. As discussed in the previous subsection, 
the emergence of an extremum in the circumferential radius $R_c$ dictates that $f_1'(0)=0$;
this leads to the result $b_1=0$. The $(\eta\eta)$ component of the Einstein equations, given in
Eq. (\ref{hheq}), yields near the throat/equator a constraint equation:
\begin{equation}\label{eq10}
\left[\left(\eta_0^2\phi'^2 + 4\right) e^{f_1} - 8 f_0'\phi' \dot{F}\right]_{\eta=0}= 0\,.
\end{equation}
The remaining three equations may then be solved to express the second-order coefficients
$(a_2, b_2, \phi_2)$ in terms of the zero and first-order coefficients in the $\eta$-expansions
(\ref{appf0})-(\ref{appff}). These are found to have the form:
\bea
    a_2&=&\frac{b_0 \left[4 b_0 \dot{F}_0^2 \phi _1^2 \left(\eta _0^2 \phi _1^2+4\right){}^2+b_0^3 \eta _0^2
   \left(\eta _0^2 \phi _1^2+4\right){}^2-128 \dot{F}_0^2 \eta _0^2 \phi _1^6 \ddot{F}_0\right]}{256
   \dot{F}_0^2 \phi _1^2 \left(b_0^2 \eta _0^2+4 \dot{F}_0^2 \phi _1^2\right)},\\
    b_2&=&-\frac{2 b_0 \phi _1^2 \ddot{F}_0}{b_0^2 \eta _0^2+4 \dot{F}_0^2 \phi _1^2},\\[1mm]
   \phi_2&=&-\frac{4 b_0 \dot{F}_0^2 \phi _1^2 \left(\eta _0^2 \phi _1^2+4\right)+b_0^3 \eta _0^2 \left(\eta _0^2
   \phi _1^2+4\right)+64 \dot{F}_0^2 \phi _1^4 \ddot{F}_0}{32 \left(b_0^2 \dot{F}_0 \eta _0^2+4
   \dot{F}_0^3 \phi _1^2\right)}.
\eea
From the above expressions, it seems that there are six free parameters in our theory: the coefficients
$(\eta_0, \;\phi_0,\; \phi_1,\;a_0,\; b_0)$ and the coupling constant $\alpha$ which is defined through the
relation  $F(\phi)=\alpha\tilde{F}(\phi)$, where $\tilde{F}$ is a dimensionless quantity. However, the
actual number of free parameters is much smaller. First of all, we notice that the field equations
(\ref{tteq})-(\ref{sceq}) are invariant under the simultaneous scaling of the coordinate $\eta$, the
constant $\eta_0$, and the scalar-field coupling constant $\alpha$,
\beq
\eta \rightarrow \lambda \eta \,, \qquad \eta_0 \rightarrow \lambda \eta_0\,, 
\qquad \alpha \rightarrow \lambda^2 \alpha\,,
\eeq
where $\lambda$ is an arbitrary constant. Therefore, we may fix $\eta_0$, which determines the scale
of the wormhole's equator/throat, to a specific value, or equivalently introduce a dimensionless
coupling parameter $\alpha/\eta_0^2$. We can also fix three of the remaining four parameters by
applying appropriate boundary conditions at infinity. Thus, by demanding asymptotic flatness, expressed
by the conditions
\beq
\lim_{\eta\to\infty}|g_{tt}|=1\\, \qquad \lim_{\eta\to\infty}g_{\eta\eta}=1\,,
\label{asym-flat}
\eeq
we may fix the $a_0$ and $b_0$ parameters, while the condition $\lim_{\eta\to\infty}\phi=\phi_\infty$
allows us to fix $\phi_1$.  Concluding, the only free parameters in the near-equator/throat area are the
dimensionless coupling constant $\alpha/\eta_0^2$ and the value $\phi_0$ of the scalar field at $\eta=0$.

Let us also examine the components of the stress-energy-momentum tensor near the
throat/equator. We find:
\bea
    &&T^t_{\;t}=\frac{1}{b_0 \eta _0^2}-\frac{4 \phi _1^2 \ddot{F}_0}{b_0^2 \eta _0^2+4 \dot{F}_0^2 \phi
   _1^2}+O\left(\eta\right),\label{Ttt-thr}\\
   &&T^\eta_{\;\eta}=-\frac{1}{b_0 \eta _0^2}+O\left(\eta \right),\label{eq15}\\
   &&T^\theta_{\;\theta}=\frac{-b_0 \eta _0^2 \phi _1^2 \ddot{F}_0 \left(\eta _0^2 \phi _1^2+4\right)+2 b_0^2 \eta
   _0^2+8 \dot{F}_0^2 \phi _1^2}{8 b_0 \dot{F}_0^2 \eta _0^2 \phi _1^2+2 b_0^3 \eta
   _0^4}+O\left(\eta\right). \label{Thh-thr}
\eea
We observe that, as desired, all components of the stress-energy tensor are finite at $\eta=0$, i.e. at the location
of the throat or equator of the solution.

\subsubsection{Expansion at large distances}
At large values of the radial coordinate, the metric functions and scalar field are expanded in
a power series form in $1/\eta$:
\bea
    e^{f_0}&=&1+\sum_{n=1}^\infty {\frac{p_n}{\eta^n}},\\
    e^{f_1}&=&1+\sum_{n=1}^\infty {\frac{q_n}{\eta^n}},\\
    \phi&=&\phi_\infty+\sum_{n=1}^\infty {\frac{d_n}{\eta^n}}\,.
\eea
In the above expressions, we have already imposed the conditions for asymptotic flatness and
constant value of the scalar field. Substituting the above expansions into the field
equations (\ref{tteq})-(\ref{sceq}), we may determine the unknown coefficients 
$(p_n, q_n, d_n)$ in terms of only two coefficients that remain arbitrary: $d_1=-D$, where $D$
is the scalar charge of the wormhole, and $p_1=-2M$, where $M$ is the Arnowitt-Deser-Misner
(ADM) mass of the wormhole. Thus, the number of free parameters at infinity is also two,
similarly to the near-throat/equator regime. We have calculated the remaining coefficients up
to order $\mathcal{O}(1/r^5)$, and the asymptotic solutions have the following form:
\bea
    e^{f_0}&=&1-\frac{2M}{\eta}+\frac{2M^2}{\eta^2}-\frac{M(D^2+36M^2-12\eta_0^2)}{24\eta^3}+\frac{D^2M^2+12(M^4-M^2\eta_0^2-4D M\dot{F}_\infty)}{12\eta^4}+\mathcal{O}\left(\frac{1}{\eta^5}\right),\label{inff0}\\
    e^{f_1}&=&1+\frac{2M}{\eta}+\frac{12M^2-D^2-4\eta_0^2}{8\eta^2}+\frac{M\left[12(M^2-3\eta_0^2)-5D^2\right]}{24\eta^3}\nonumber\\
    &&+\frac{3D^4+D^2(96\eta_0^2-104M^2)+48(M^4-24M^2\eta_0^2+7\eta_0^4)+1536D M\dot{F}_\infty}{768\eta^4}+\mathcal{O}\left(\frac{1}{\eta^5}\right),\label{inff1}\\
    \phi&=&\phi_\infty-\frac{D}{\eta}-\frac{D^3+4D(M^2-3\eta_0^2)}{48\eta^3}-\frac{4M^2\dot{F}_\infty}{\eta^4}+\mathcal{O}\left(\frac{1}{\eta^5}\right)\label{infff}
\eea
We observe that the above solutions have exactly the same form as the corresponding solutions
which describe asymptotically-flat black holes \cite{Antoniou:2017acq}. Apparently, the emergence
of an asymptotically-flat limit does not depend on the choice of the boundary condition at the
other asymptotic regime i.e. the horizon of a black hole or the throat/equator of a wormhole.
The main difference is that, in the case of black holes, the mass $M$ and the scalar charge $D$
are related parameters -- which makes black holes a one-parameter family of solutions -- while,
in the case of wormholes, these two parameters are independent.  Also, the aforementioned asymptotic
solutions at infinity are almost independent of the functional form of the coupling function
$F(\phi)$, since the latter does not enter in the expansions earlier than in the fourth order. 

Finally, if we make use of the expansions above, we may calculate again the stress-energy tensor
components at large distances. These are found to be 
\begin{equation}
    T^t_{\;t}=-T^\eta_{\;\eta}=T^\theta_{\;\theta}=T^\varphi_{\;\varphi}\approx -\phi'^2/4 \approx -D^2/4\eta^4+\mathcal{O}\left(1/\eta^5\right). \label{Tmn-far}
\end{equation}
As we expect, the above expressions have exactly the same form as the corresponding ones for the asymptotically
flat black holes. We observe that, at large distances where the curvature of spacetime is small, the stress-energy
tensor is dominated by the kinetic term of the scalar field which is itself decaying fast.


\subsection{Violation of Energy Conditions}
In our previous works \cite{Kanti:2011jz,Kanti:2011yv}, we have shown that for any single-throat wormhole
the null energy condition is violated at least in some region near the throat. Here, we
review that analysis, and show that the violation of the null energy condition also
holds for double-throat wormholes.

The null energy condition (NEC) is expressed as $ T_{\mu\nu} n^\mu n^\nu \geq 0$, where
$n^\mu$ is any null vector satisfying the condition $n^\mu n_\mu=0$. We may define the
null vector as $n^\mu=\left(1,\sqrt{-g_{tt}/g_{\eta\eta}},0,0\right)$ with its contravariant
form being $n_\mu=\left(g_{tt},\sqrt{-g_{tt}\,g_{\eta\eta}},0,0\right)$. For a
spherically-symmetric spacetime, the NEC takes the form:
\beq
T_{\mu\nu}n^\mu n^\nu=T^t_t n^t n_t + T^\eta_\eta n^\eta n_\eta
=-g_{tt}\,(-T^t_t +T^\eta_\eta)\,.
\eeq
Then, the NEC holds if $ -T_t^t + T_\eta^\eta \geq 0$. Alternatively, we may choose
$n^\mu=\left(1,0,\sqrt{-g_{tt}/g_{\theta \theta}},0\right)$, and a similar analysis
leads to the condition $-T_t^t + T_\theta^\theta \geq 0 $. 

For a wormhole solution to emerge, it is essential that these two conditions
are violated \cite{Morris:1988cz}. Indeed, using the expansion of the wormhole solution at the
throat/equator, we find 
\beq
\left[ -T_t^t + T_\eta^\eta \right]_{\eta_{t,e}} 
  = -2 \left[e^{-f_1} R_c''/R_c \right]_{\eta_{t,e}}\,.
  \eeq
Consequently, the NEC is always violated at the throat(s), 
since $R_c$ possesses a minimum there, implying $R_c''(\eta_{t}) > 0$, while
no violation occurs at the equator, where $R_c''(\eta_{e}) < 0$. For example,
for a single-throat solution with the throat at $\eta=0$, we obtain the
explicit expressions
\begin{eqnarray}
\left[ -T_t^t + T_\eta^\eta \right]_{\eta=0} &  = & 
\left[ -\frac{2 e^{-f_1}}{\eta_0^2}
      +\frac{4 \ddot{F}\phi'^2}{e^{2f_1}\eta_0^2 +4 \dot{F}^2\phi'^2}\right]_{\eta=0}=
      -\frac{2}{b_0 \eta_0^2} + \frac{4 \phi _1^2 \ddot{F}_0}{b_0^2 \eta _0^2+4 \dot{F}_0^2 \phi _1^2}\,, 
\label{nec1} \\
\left[ -T_t^t + T_\theta^\theta \right]_{\eta=0} &  = & 
\left[\frac{\ddot{F}\phi'^2\left(4-\eta_0^2\phi'^2\right)}
       {2\left(e^{2f_1}\eta_0^2 +4 \dot{F}^2\phi'^2\right)}\right]_{\eta=0} 
       =\frac{\phi _1^2 \ddot{F}_0 \left(4-\eta _0^2 \phi _1^2\right)}
       {2 b_0^2 \eta _0^2+8 \dot{F}_0^2 \phi_1^2}\,,
\label{nec2}   
\end{eqnarray}
where we have used the approximate expressions Eqs. (\ref{Ttt-thr})-(\ref{Thh-thr}) near the wormhole
throat. We note that the desired violation of the NEC follows not from the presence of an
exotic form of matter but from the synergy between the scalar field and the quadratic GB term.

In the far-asymptotic regime, we may use the expansions at infinity Eqs. (\ref{inff0})-(\ref{infff})
to find that the two Null Energy Conditions take the form:
\bea
    -T^t_{\;t}+T^\eta_{\;\eta}&=&\,\frac{D^2}{2\eta^4}+\mathcal{O}\left(\frac{1}{\eta^5}\right),\\
    -T^t_{\;t}+T^\theta_{\;\theta}&=&-\frac{40DM\dot{F}_\infty}{\eta^6}+\mathcal{O}\left(\frac{1}{\eta^7}\right).
\eea
We observe that if $D\dot{F}_\infty>0$, the second Null Energy Condition is also violated at spatial infinity.

Let us also examine the Weak Energy Condition (WEC), which suggests that the energy density measured by
any observer has to be greater than or equal to zero. This is expressed through the inequality:
$T_{\mu\nu}V^\mu V^\nu \ge 0$,
where $V^\mu$ is any timelike vector. If we choose $V^\mu=(1/\sqrt{-g_{tt}},0,0,0)$, 
and impose the condition $V_\mu V^\mu = -1$, then $V_\mu=(-\sqrt{-g_{tt}},0,0,0)$, and the WEC is simply
$T^t_{\;t}\le 0$. Near the throat/equator, we found that $T^t_{\;t}$ is given by Eq. (\ref{Ttt-thr});
this expression is not sign-definite, therefore the WEC may also be violated in the small $\eta$-regime. 
On the other hand, at asymptotic infinity, where $T_{\mu\nu}$ is dominated by the kinetic term of the
scalar field, the $T^t_{\;t}$ component is given by Eq. (\ref{Tmn-far}) and clearly obeys the WEC.


\section{Numerical solutions}
We now turn to the derivation of the wormhole solutions by numerically integrating
the three second-order, ordinary differential equations (\ref{tteq}), (\ref{ffeq})
and (\ref{sceq}). In order to find asymptotically-flat, regular wormhole solutions,
we have to impose appropriate boundary conditions at asymptotic infinity and at the
throat/equator, as discussed in the previous section. For completeness, we list
here the full set of these boundary conditions:
\begin{eqnarray}
& & f_0(\infty) = f_1(\infty) = 0 \ , \ \ \phi(\infty) = \phi_\infty \ , 
\label{bcinfty}\\
& & f_1'(0) = 0\ , \ \ 
\left[\left(\eta_0^2\phi'^2 + 4\right) e^{f_1} - 8 f_0'\phi' \dot{F}\right]_{\eta=0}= 0 \ . 
\label{bcs} 
\end{eqnarray}
For the numerical integration, we use the compactified coordinate $x=\eta/(\eta+\eta_0)$
to cover the range $0\leq \eta < \infty$. 
We choose $\eta_0=1$ for all our numerical solutions.
The software package COLSYS is then used to
solve the three, second-order ODEs with the aforementioned boundary conditions.

In our analysis, we have found wormhole solutions with either vanishing or non-vanishing
asymptotic values of the scalar field, namely for $\phi_\infty=0$ and $\phi_\infty=1$.
We have also considered several forms of $F(\phi)$, including exponential $F=\alpha e^{-\gamma\phi}$,
$F=\alpha e^{-\gamma\phi^2}$, power-law $F=\alpha \phi^n$ with $n\neq 0$, inverse power-law
$F=\alpha \phi^{-n}$, and logarithmic  $F=\alpha \ln(\phi)$ functions. We have found wormhole
solutions in every single case studied. Due to the qualitative similarity of the obtained
behaviour for the metric functions and scalar field, in this work we will mainly focus on the
presentation of results for the cases with coupling functions $F=\alpha e^{-\phi}$ and 
$F=\alpha \phi^2$, and present combined graphs for different forms of $F(\phi)$ whenever
possible. During our quest for regular, physically-acceptable wormhole solutions,
spontaneously scalarized black holes also emerged in multitude thus confirming the
results of \cite{Antoniou:2017acq}.

%
\begin{figure}[t]
\begin{center}
(a)\mbox{\includegraphics[width=.47\textwidth, angle =0]{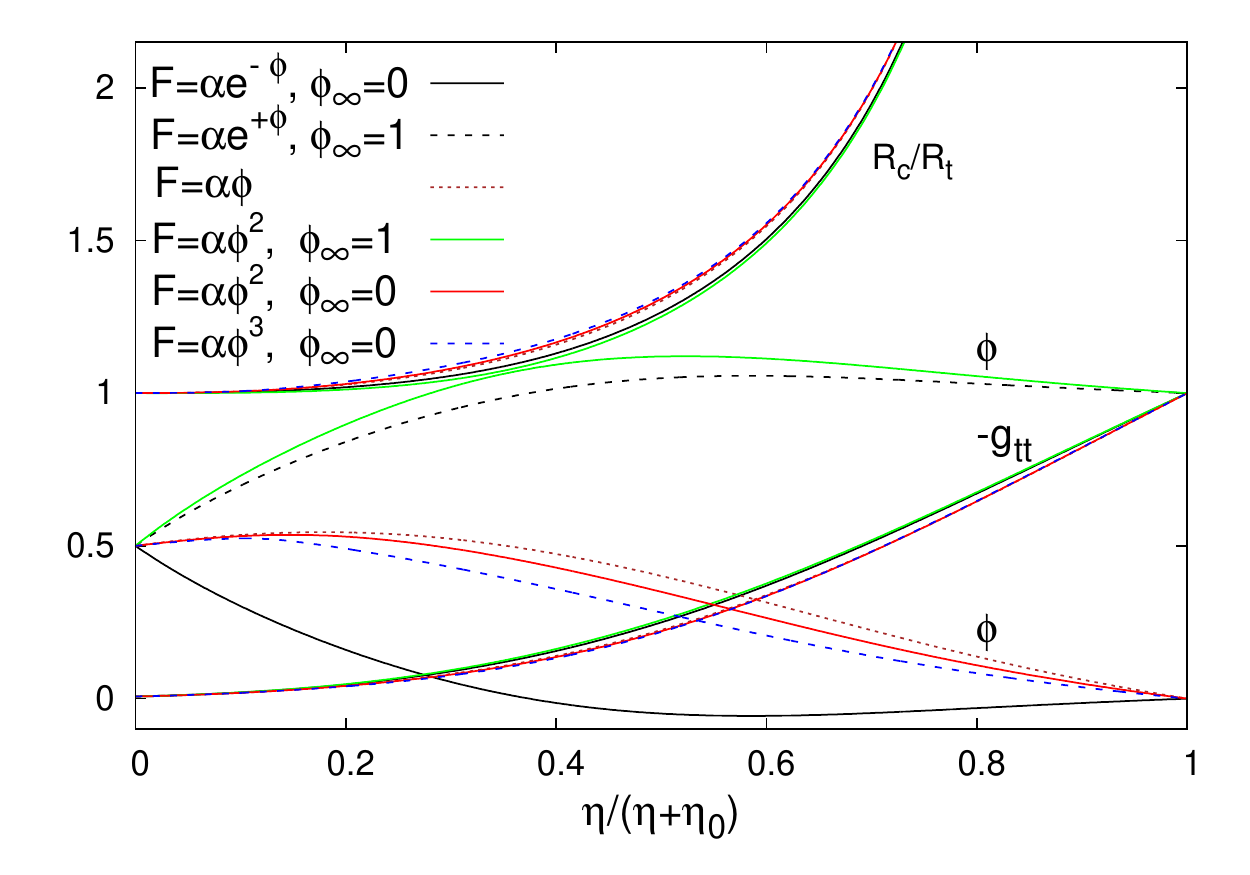}}
(b)\mbox{\includegraphics[width=.47\textwidth, angle =0]{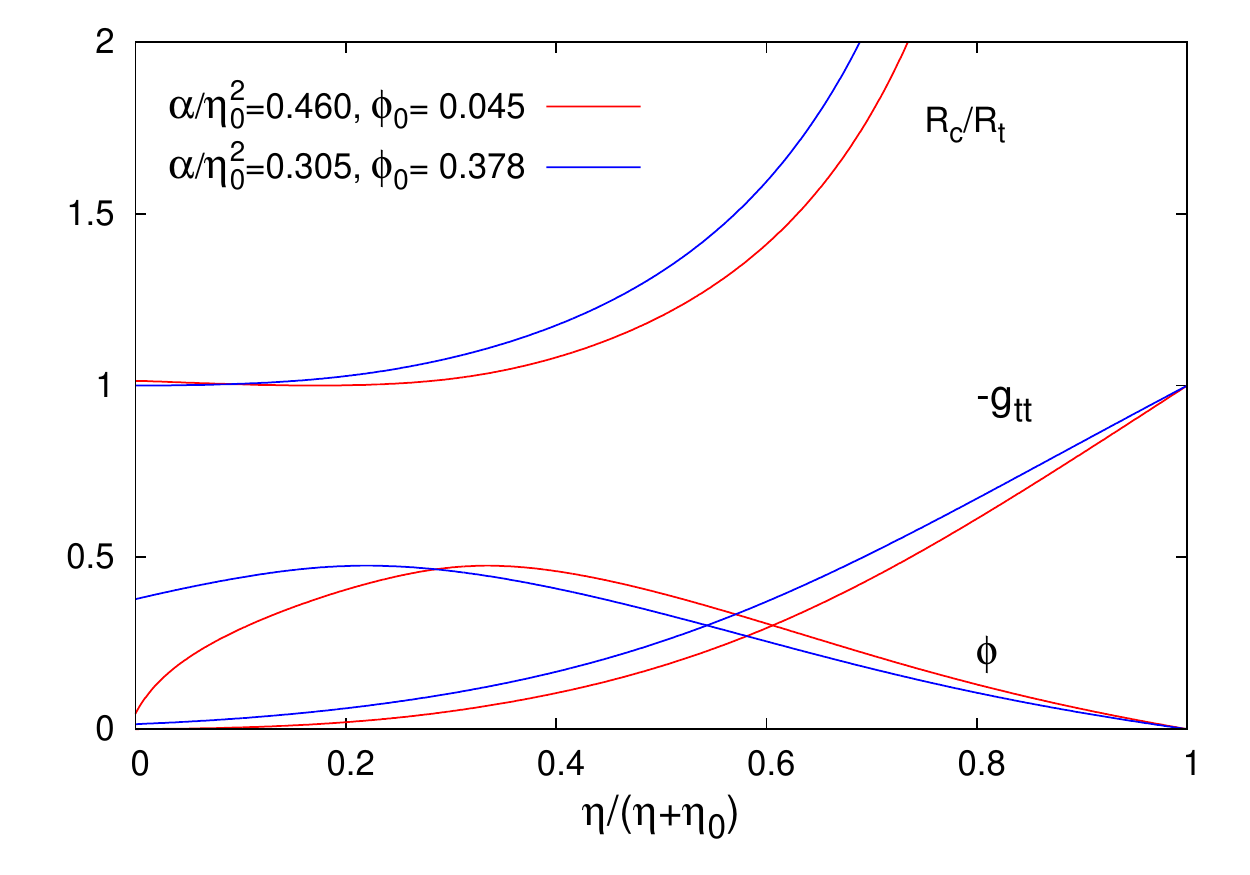}}
\end{center}
\caption{Solutions:
(a) The metric component $-g_{tt}$, the scalar field $\phi$ and the scaled circumferential radius
$R_c/R_t$ are shown as functions of the compactified coordinate $\eta/(1+\eta)$ for different
coupling functions. All solutions are characterized by the same values of $f_0(0)$ and $\phi(0)$.
(b) The metric component $-g_{tt}$, the scalar field $\phi$ and the scaled circumferential radius
$R_c/R_t$ are shown as function of the compactified coordinate $\eta/(1+\eta)$ for
a single throat wormhole (blue) and a double throat wormhole (red) for the same values 
of the scaled scalar charge and the scaled throat area.
\label{fig1}
}
\end{figure}

In Fig.~\ref{fig1}(a), we depict the metric component $-g_{tt}=e^{f_0}$, the scalar field $\phi$ 
and the scaled circumferential radius $R_c/R_t$ for several coupling functions $F(\phi)$.
All solutions are characterised by the same boundary values $f_0(0)=-5$ and
$\phi(0)=0.5$ but we have allowed for two different asymptotic values for $\phi$, namely
$\phi_\infty=0$ and $\phi_\infty=1$. We observe that the behaviour of the metric component
$-g_{tt}$ and the circumferential radius $R_c/R_t$ depends rather mildly on the form
of the coupling function or the asymptotic value $\phi_\infty$. 
On the other hand, both of these factors considerably affect the profile of the scalar field
as may be clearly seen from the plot.

We have found both single and double-throat wormhole solutions for every form of the
coupling function $F(\phi)$. In  Fig.~\ref{fig1}(b), we compare single and double-throat
wormholes for the same values of the scaled scalar charge $D/M$ and scaled throat area
$A_t/16\pi M^2$. Once again, it is the scalar field that is mostly affected by the
different geometry near the throat or equator. We note for future reference that the
derivatives of the $-g_{tt}$ and $\phi$ do not vanish at $\eta=0$, i.e. at the throat,
for single-throat wormholes, or at the equator, for double-throat ones. This feature
will lead to the introduction of a distribution of matter, albeit a physically-acceptable
one, at $\eta=0$ when we attempt to symmetrically continue our wormhole solutions to the
negative regime of the $\eta$ coordinate. This process and the implications of the
associated junction conditions will be studied in Section IV. 

%
\begin{figure}[t!]
\begin{center}
(a)\mbox{\includegraphics[width=.46\textwidth, angle =0]{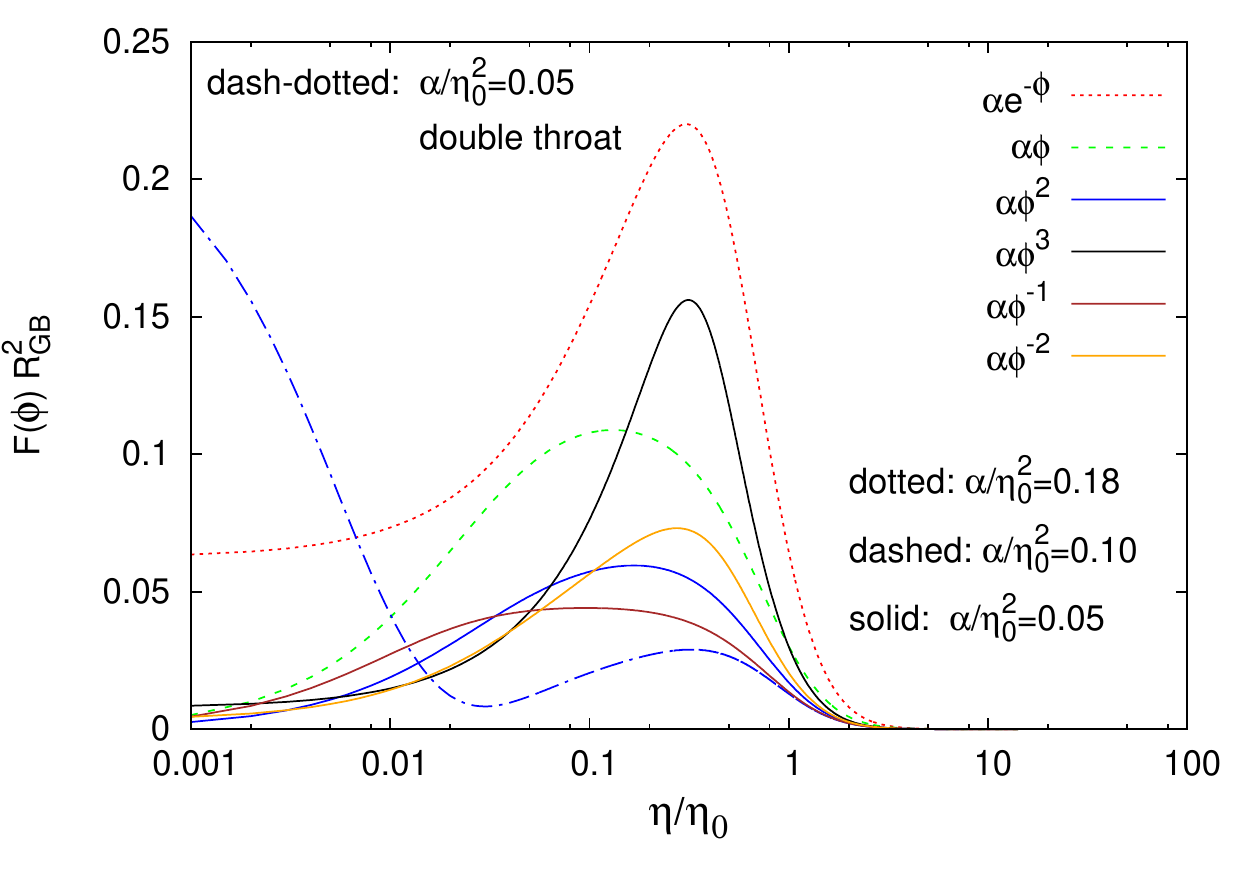}
(b)       \includegraphics[width=.46\textwidth, angle =0]{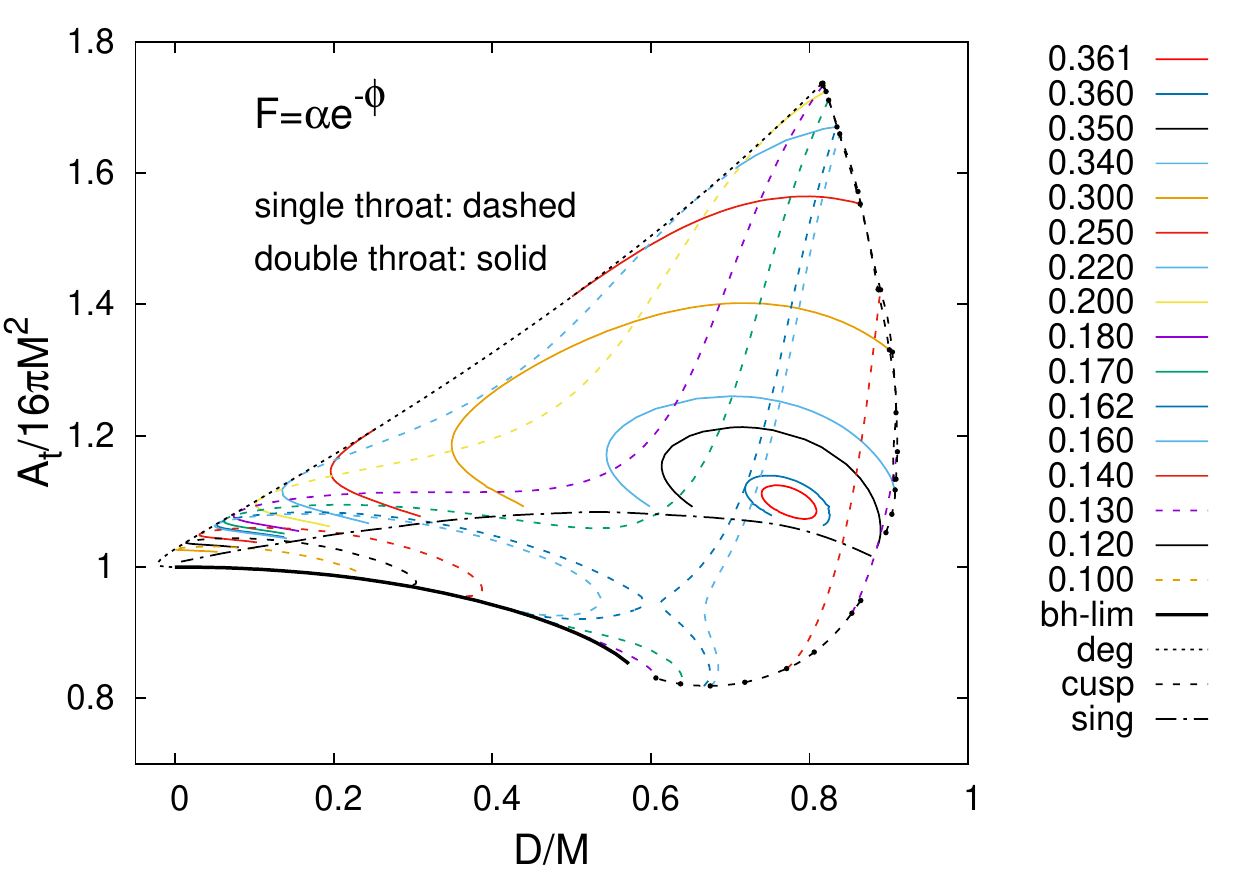}}
\end{center}
\caption{
(a) The quantity $F(\phi) R^2_{GB}$ as function of $\eta$ for several forms of the coupling
function $F(\phi)$.  
(b) Domain of existence: The scaled throat area of single and double throat wormholes
is shown as function of the scaled scalar charge for the coupling function
$F=\alpha e^{-\phi}$ for several values of $\alpha/\eta_0^2$. 
\label{fig2}
}
\end{figure}

The spacetime around our wormhole solutions is finite for all values of the radial
coordinate $\eta \in [\,0, \infty)$. All curvature invariant quantities remain 
everywhere finite, as expected. In Fig. \ref{fig2}(a), we depict the profile of the
quantity $F(\phi) R^2_{GB}$, for a variety of forms of the coupling function $F(\phi)$ and
for the same set of values of the free parameters for easy comparison. We observe
that the combination $F(\phi) R^2_{GB}$ is indeed finite, vanishes at asymptotic infinity
as anticipated while its profile in the small $\eta$-regime depends on the form of $F(\phi)$.
We also note that the double-throat solution presents a different profile from the 
single-throat ones; this is due to the fact that the value of the scalar field at the
equator is different from its value at the throat.

%
\begin{figure}[b!]
\begin{center}
(a)\mbox{\includegraphics[width=.46\textwidth, angle =0]{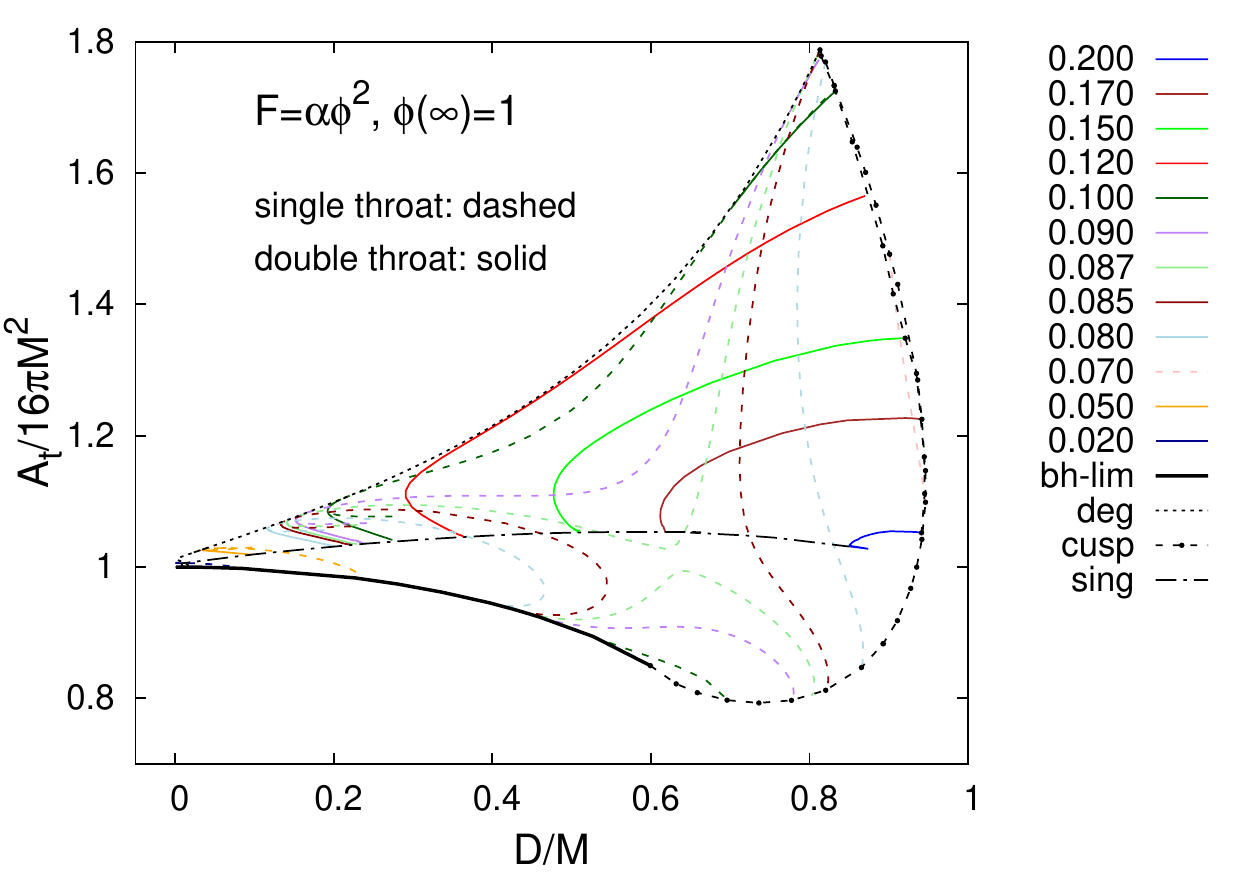}
(b)       \includegraphics[width=.46\textwidth, angle =0]{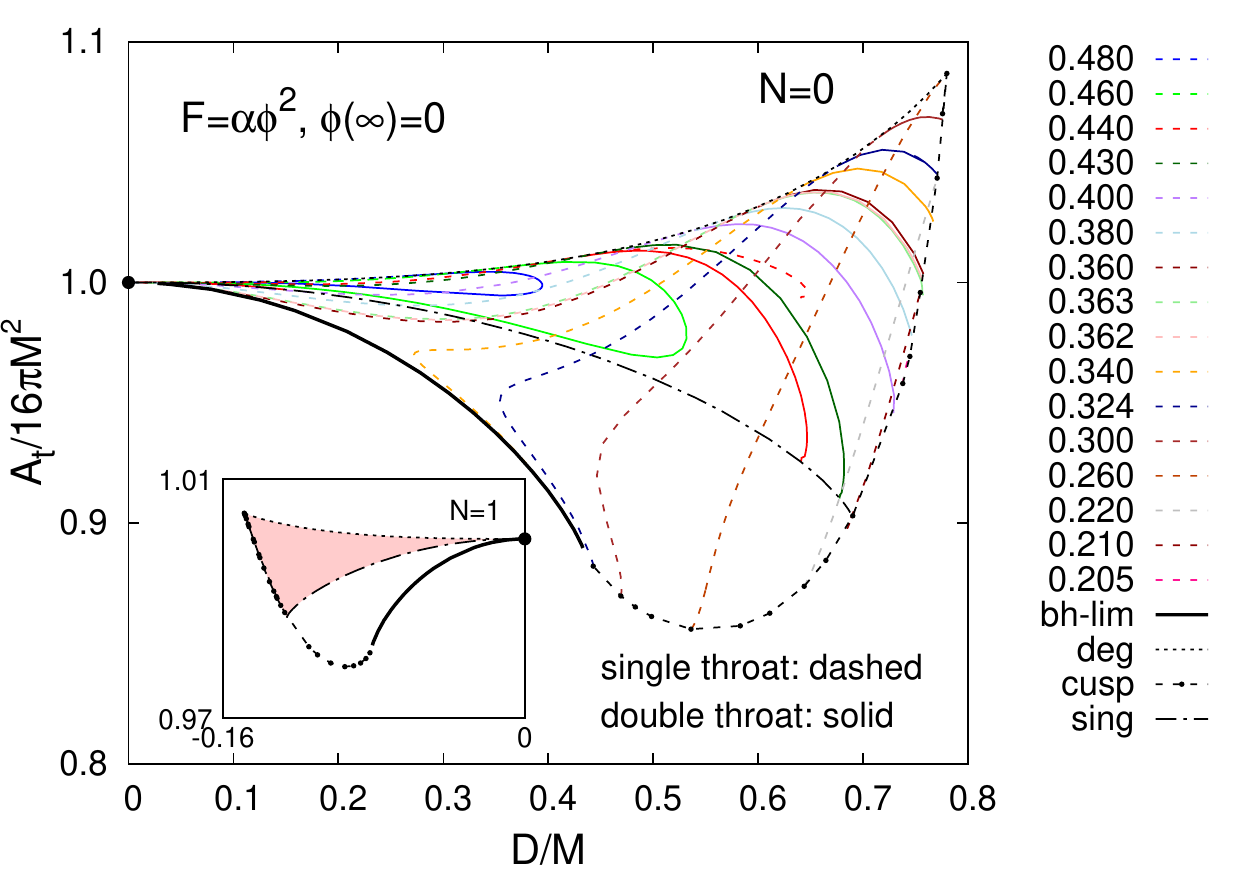}}
\end{center}
\caption{Domain of existence for the coupling function $F=\alpha \phi^2$ for several
values of $\alpha/\eta_0^2$ : 
The scaled throat area of single and double throat wormholes is shown as function of
the scaled scalar charge for (a) $\phi_\infty =1$, and (b) $\phi_\infty =0$.
The dot indicates the Schwarzschild black hole. The inlet shows the domain of existence
for wormholes with one node of the scalar field -- the red area indicates the domain
where single and double-throat wormholes co-exist.
\label{fig3}
}
\end{figure}

Next, we discuss the domain of existence (DOE) of the wormhole solutions,
in terms of the scaled scalar charge and the scaled throat area, and restrict our
discussion to the indicative cases of the exponential and quadratic coupling functions.
In Fig.~\ref{fig2}(b), we show the DOE for the exponential case,
$F=\alpha e^{-\phi}$. The different curves correspond to families of wormholes
for a fixed value of $\alpha$ with single throat (dashed) and double throat (solid). 
Solutions emerge for arbitrarily small values of $\alpha$
up to some maximal value - here, we depict a variety of solutions arising up to
the value $\alpha/\eta_0^2=0.361$.
The boundary of the DOE is formed by the black hole solution with
scalar hair (solid black), the wormhole solutions with a degenerate throat (dotted black),
configurations with cusp singularities outside the throat (dashed black) and configurations
with singularities at the equator (dashed-dotted black). We note that the part of the
DOE above the dashed-dotted curve comprises both single-throat and
double-throat wormholes. The single-throat wormholes of this area can in fact be
obtained from the double-throat ones -- we will return to this point in Section IV.
The region of the domain of existence below the dashed-dotted curve contains only
single-throat wormholes which are not related to double-throat solutions.

We now turn to the case of the quadratic coupling function, $F=\alpha \phi^2$. 
Contrary to what happens in the case of the exponential coupling function, in this case,
the DOE depends on the asymptotic value of the scalar field. For $\phi(\infty)=1$,
the quantity $\dot{F}(\phi)$ assumes a non-zero asymptotic value, as in the exponential case,
therefore the DOE, depicted in Fig.~\ref{fig3}(a), is similar to the one displayed in Fig. \ref{fig2}(b).
In contrast, if $\phi(\infty)=0$, then $\dot{F}$ vanishes asymptotically and the range
of $\alpha$, for which wormholes arise, is also limited from below. The DOE
in this case is shown in Fig.~\ref{fig3}(b) -- now, wormholes emerge only
if $0.205 < \alpha/\eta_0^2 <0.480$. The Schwarzschild black holes are now part of
the boundary of the DOE, as indicated by the dot in Fig.~\ref{fig3}(b),
since the constant configuration $\phi \equiv \phi_\infty=0$ solves the scalar field equation trivially.
Moreover, wormhole solutions exist for which the scalar field may possess $N$ nodes.
The boundary of the DOE for $N=1$ is shown in the inlet in Fig.~\ref{fig3}(b).
Note that the range of $\alpha$ in this case is approximately $1.85\leq \alpha/\eta_0^2 \leq 2.75$, 
i.~e.~considerably larger than for $N=0$.

\begin{figure}[t!]
\begin{center}
(a)\mbox{\includegraphics[width=.46\textwidth, angle =0]{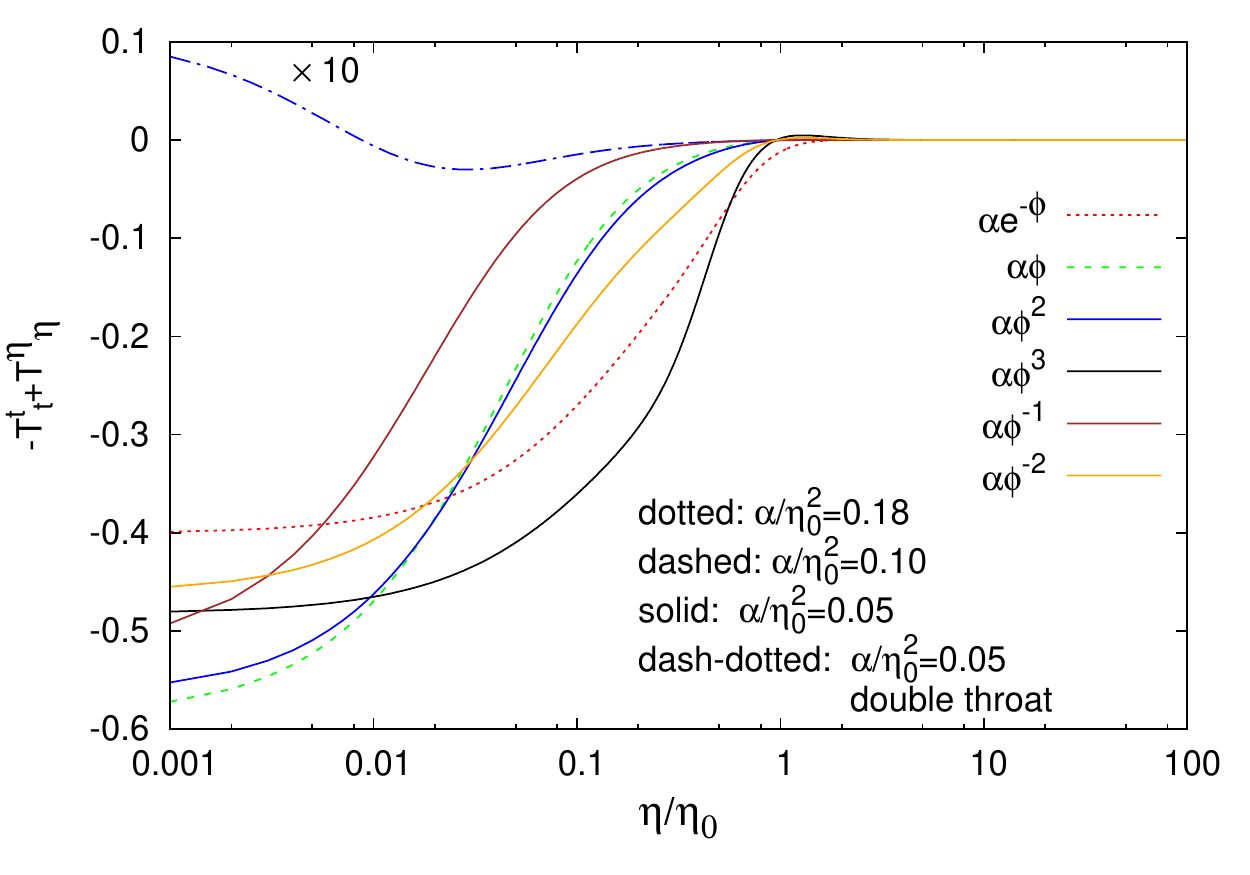}
(b)       \includegraphics[width=.46\textwidth, angle =0]{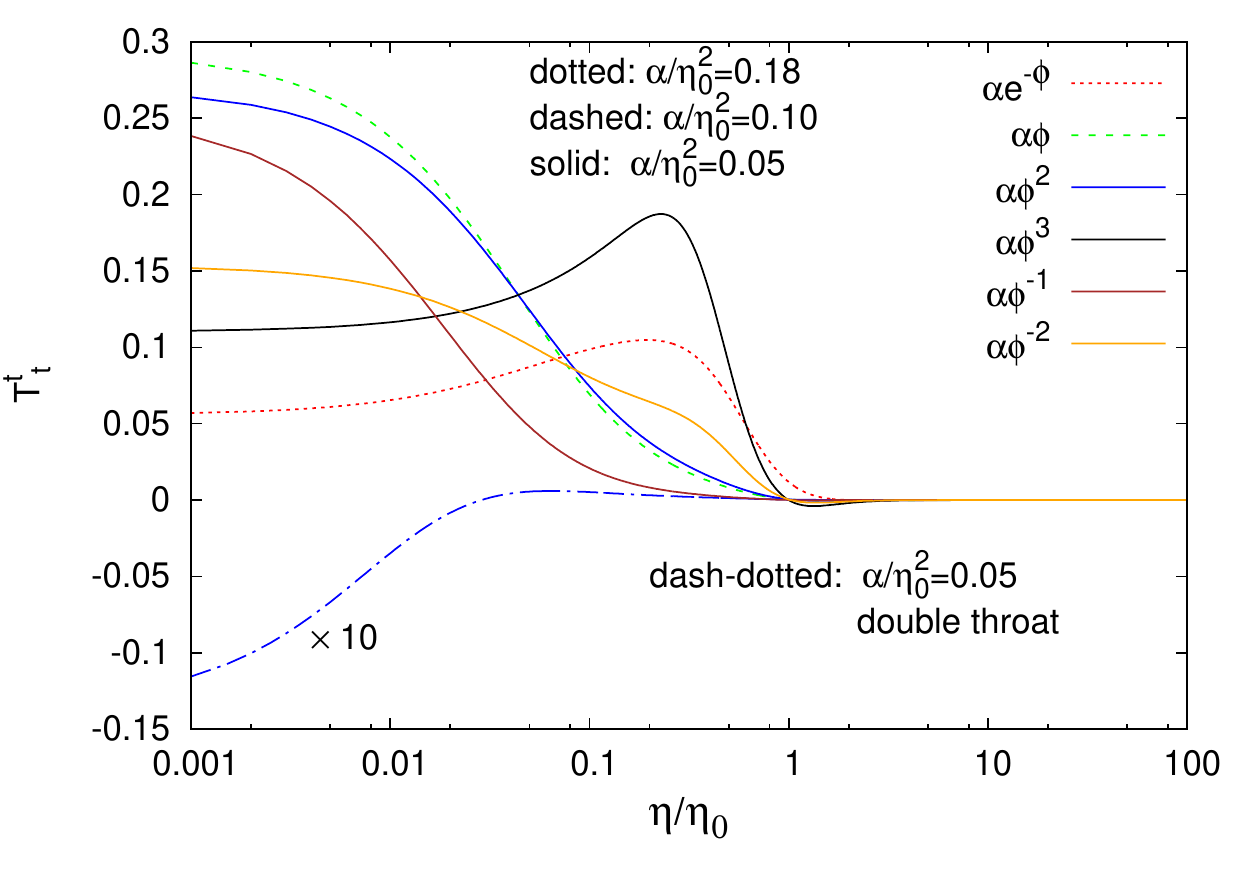}}
\end{center}
\caption{(a) The Null Energy Condition and (b) the Weak Energy Condition for
a variety of forms of the coupling function $F(\phi)$.}
\label{fig4} 
\end{figure}

Let us finally address the issue of the violation of the Null and Weak Energy Conditions. In 
Fig.~\ref{fig4}(a), we display the quantity $-T_t^t + T_\eta^\eta$ for a number of wormhole
solutions arising for different forms of the coupling function $F(\phi)$.  It is evident that
the NEC is always violated near the throat of each solution by an amount which depends
on the form of the coupling function $F(\phi)$ since the latter determines the weight of the
GB term in the theory. On the other hand, the NEC is obeyed at asymptotic infinity. We note
that, in the case of the double-throat solution, the NEC is violated at the throat while it is
obeyed at the equator, according to the analysis of the previous section. A similar behaviour
is exhibited by the $T_t^t$ component depicted in Fig. ~\ref{fig4}(b): the WEC is again violated
at the small $\eta$-regime, by an amount determined by $F(\phi)$, while it is obeyed at
asymptotic infinity where the GB term becomes negligible. The double-throat solution 
again respects the WEC at the equator while it violates it near the throat.

\section{Junction conditions} 

\begin{figure}[t!]
\begin{center}
\includegraphics[width=.52\textwidth, angle =0]{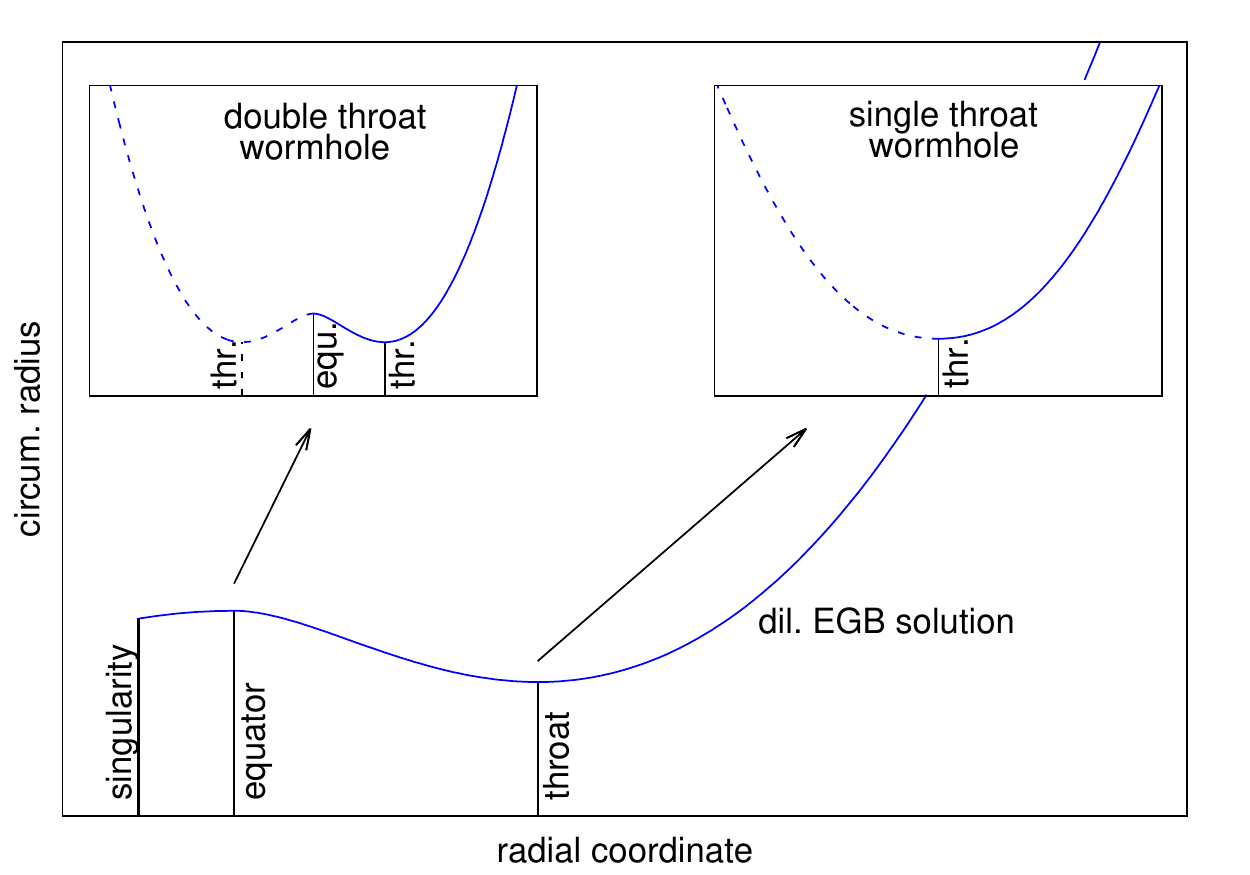}
\caption{Schematic picture for the construction of double-throat and single-throat
wormhole solutions.}
\end{center}
\label{fig5} 
\end{figure} 

Wormhole solutions may be either symmetric
or asymmetric under the change $\eta \rightarrow - \eta$. In the context of the EsGB theory
with an exponential coupling function \cite{Kanti:2011jz,Kanti:2011yv}, asymmetric wormholes were found
but they were plagued by curvature singularities lurking behind the throat. A regular
wormhole solution may then be constructed by imposing a symmetry under the change
$\eta \rightarrow - \eta$. The obtained solution then consists of two parts: the first
coincides with the part of the asymmetric solution which extends from the asymptotic
region at infinity to the location of the throat; the second part of the wormhole
solution is obtained by the symmetric continuation of the first part in the negative
$\eta$-regime.

A similar construction was performed in the context of the present analysis, in
the case of solutions with a single throat -- these solutions are the ones depicted in
Figs. \ref{fig2}(b) and \ref{fig3}(a,b) under the dashed-dotted curves. In the case
of singular wormhole solutions with a throat and an equator, a similar process may
give rise to double-throat wormholes and to single-throat wormholes since
now there are two options, as Fig. 5 depicts. The first option is to 
construct a regular wormhole by cutting at the throat and symmetrically continuing
to the left, as described above; in that case, the equator is removed from the
spacetime geometry and a single-throat wormhole is constructed. The second option
is to cut the singular solution at the equator, keep the regular part from the
asymptotic infinity to the equator and continue symmetrically to the left; in
this way, a double-throat wormhole solution, with an equator located exactly
between the throats, is constructed. Both wormholes possess the same mass and
scalar charge, since these quantities are extracted from the asymptotic region
that is common in both solutions. Hence, for any double-throat wormhole there
exist a single-throat wormhole with the same mass and scalar charge -- these are
the solutions depicted in Figs. \ref{fig2}(b) and \ref{fig3}(a,b) above the
dashed-dotted curves.

Let us now discuss in more detail the construction of symmetric, regular, and
thus traversable, wormholes
\footnote{As the coupling function may acquire different forms in the context of our analysis,
constructing also asymmetric wormholes, see e.g. \cite{Bronnikov:2017kvq}, may be indeed a possibility
that we plan to pursue in a future work.}.  
From Fig. 1, we observe that the derivatives of the $-g_{tt}$ and $\phi$ do not vanish
in general at $\eta=0$. Therefore, imposing a symmetry under
$\eta \rightarrow - \eta$ creates a ``cusp'' in the profile of the aforementioned quantities.
This feature may be attributed to the presence of a distribution of matter at $\eta=0$,
i.e. around the throat or the equator, for single or double-throat solutions, respectively. 
The embedding of this thin-shell matter distribution in the context of the complete
solution is determined through the junction conditions \cite{Israel:1966rt,Davis:2002gn}, that follow by
considering the jumps in the Einstein and scalar field equations (\ref{einfldeq}) as
$\eta \rightarrow -\eta$. These are found to have the form
\begin{equation}
\langle G^\mu_{\phantom{a}\nu} -T^\mu_{\phantom{a}\nu}\rangle = s^\mu_{\phantom{a}\nu} \ , \ \ \ 
\langle \nabla^2 \phi + \dot{F} R^2_{\rm GB}\rangle = s_{\rm scal} \ ,
\label{jumps}
\end{equation}
where $s^\mu_\nu$ denotes the stress-energy tensor of the matter
at the throat, resp. equator, and $s_{\rm scal}$ a source term for the scalar field.
For a physically-acceptable solution, this matter distribution should not be exotic.
We thus assume a perfect fluid with pressure $p$ and energy density $\rho$, and
a scalar charge $\rho_{\rm scal}$ at the throat, resp. equator, together with the 
gravitational source \cite{Kanti:2011jz,Kanti:2011yv}
\begin{equation}
S_\Sigma = \int \left[\lambda_1 + 2 \lambda_0 F(\phi) \bar{R}\right]\sqrt{-\bar{h}} d^3 x
\label{act_th}
\end{equation}
where $\lambda_1,\lambda_0$ are constants, $\bar{h}_{ab}$ is the three-dimensional
induced metric at the throat, resp. equator, and $\bar{R}$ is the corresponding Ricci
scalar. Substitution of the metric then yields the junction conditions
\begin{eqnarray}
8 \dot{F} \phi' e^{-\frac{3 f_1}{2}}
& = &
\lambda_1\eta_0^2 + 4\lambda_0 F  e^{-f_1}- \rho\eta_0^2 \ , 
\label{j_00}\\
e^{-\frac{f_1}{2}} f_0' 
& = &
\lambda_1 + p  \ , 
\label{j_tt}\\
e^{-f_1}\phi' - 4  \frac{\dot{F}}{\eta_0^2} f_0' e^{-2 f_1}
& = &
-4\lambda_0\frac{\dot{F}}{\eta_0^2} e^{-\frac{3 f_1}{2}} +\frac{\rho_{\rm scal}}{2} \ ,
\label{j_ss}
\end{eqnarray}
%
%
%
where all quantities are taken at $\eta=0$. 
The above junction conditions determine $\rho$, $p$ and $\rho_{scal}$
in terms of the arbitrary constants $\lambda_0$ and $\lambda_1$ and the form of the scalar
field and metric functions close to the boundary. 

\begin{figure}[t]
\begin{center}
\mbox{\includegraphics[width=.47\textwidth, angle =0]{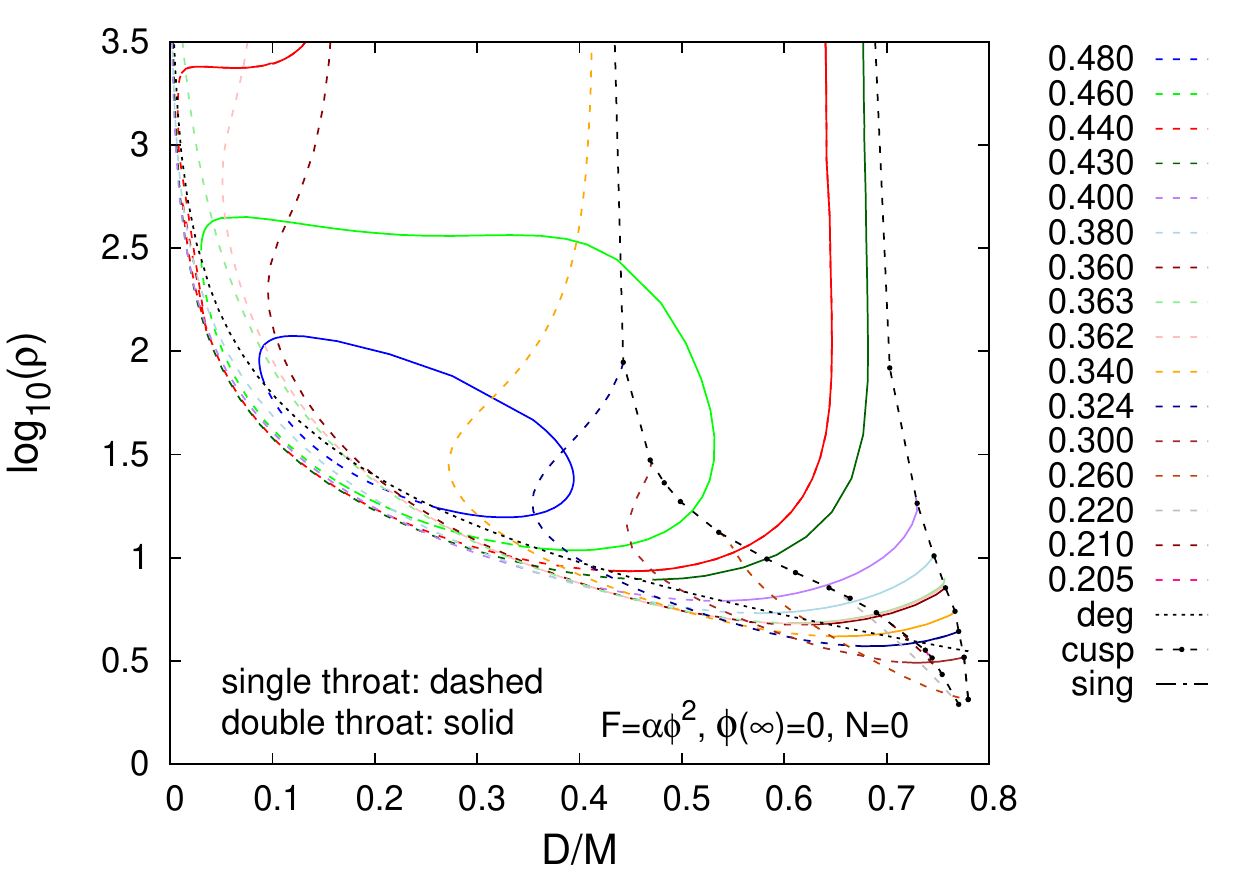}
}
\end{center}
\caption{
The energy density $\rho$ at $\eta=0$, for $F(\phi)=\alpha \phi^2$ for several values of
$\alpha/\eta_0^2$, $p=0$ (dust) and specific values of $(\lambda_0, \lambda_1)$.
\label{fig6}
}
\end{figure}

For every form of the coupling function $F(\phi)$, we may find an extensive
($\lambda_0$, $\lambda_1$)-parameter regime over which $\rho$ is always positive, and the
necessity of the exotic matter is thus avoided. An interesting special case is when the
matter distribution around the throat has a vanishing pressure, i.e. $p=0$, and therefore
its equation of state is the one of dust. In this case, Eq. (\ref{j_tt}) gives
$\lambda_1=e^{-f_1/2}f_0'$. If we choose $\lambda_0=\lambda_1$, Eqs. (\ref{j_00}) and
(\ref{j_ss}) easily yield
\beq
    \rho=\frac{e^{-\frac{3 f_1}{2}}}{\eta_0}\left[  \left(  4 F + \eta_0^2 e^{f_1} \right)f_0'
    -8\dot F \phi'     \right], \qquad 
    \rho_{\rm sc}=2 e^{-f_1} \phi',\label{rhomat} 
\eeq
respectively, where again all quantities are evaluated at $\eta=0$.
In Fig. \ref{fig6}, we depict the energy density $\rho$ at the throat, resp. equator,
as a function of the scaled scalar charge $D/M$, for a variety of wormhole
solutions arising for $F(\phi)=\alpha \phi^2$ and for
the aforementioned values of $p$, $\lambda_0$ and $\lambda_1$. We note that in this
example the energy density $\rho$ is positive for all wormhole solutions.
As in the construction of the solution, where the synergy of an ordinary distribution
of matter with a gravitational source kept the throat, resp. equator, open, here
a similar synergy creates a symmetric wormhole free of singularities.


\section{Embedding Diagram}

A useful way to visualize the geometry of a given manifold is the construction of the
corresponding embedding diagram. In this case, we consider the isometric embedding of
the equatorial plane of our wormhole solutions, defined as the line-element (\ref{metric})
for $t=const.$ and $\theta=\pi/2$. The isometric embedding follows by equating the
line-element of the two-dimensional equatorial plane with a hypersurface in the
three-dimensional, Euclidean space, namely 
\begin{equation}
e^{f_1}\,[d\eta^2 +(\eta^2+\eta_0^2)\,d\varphi^2]=
dz^2 +dw^2 +w^2 d\varphi^2\,, \label{emb}
\end{equation}
where ($z$, $w$, $\varphi$) is a set of cylindrical coordinates on the hypersurface.
Considering $z$ and $w$ as functions of $\eta$, we find
\bea
    w&=&e^{f_1/2}\sqrt{\eta^2+\eta_0^2}, \label{wdef}\\[2mm]
    \left(\frac{dw}{d\eta}\right)^2+\left(\frac{dz}{d\eta}\right)^2&=&e^{f_1}\,.\label{zzde}
\eea
Then, combining the above equations, we find
\beq
   z(\eta)=\pm \int_0^\eta \sqrt{e^{f_1(\tilde \eta)} -\left( \frac{d}{d\tilde \eta}
   \left[  e^{f_1(\tilde \eta)/2}\sqrt{\tilde \eta^2+\eta_0^2} \right]  \right)^2}d\tilde \eta. \label{zeq}
\eeq 
Therefore, $\{w(\eta), z(\eta)\}$ is a parametric representation of a slice of the embedded
$\theta = \pi/2$-plane for a fixed value of the $\varphi$ coordinate, while the corresponding
surface of revolution is the three-dimensional representation of the wormhole's geometry. 

\begin{figure}[t]
\begin{center}
(a)\mbox{\includegraphics[width=.46\textwidth, angle =0]{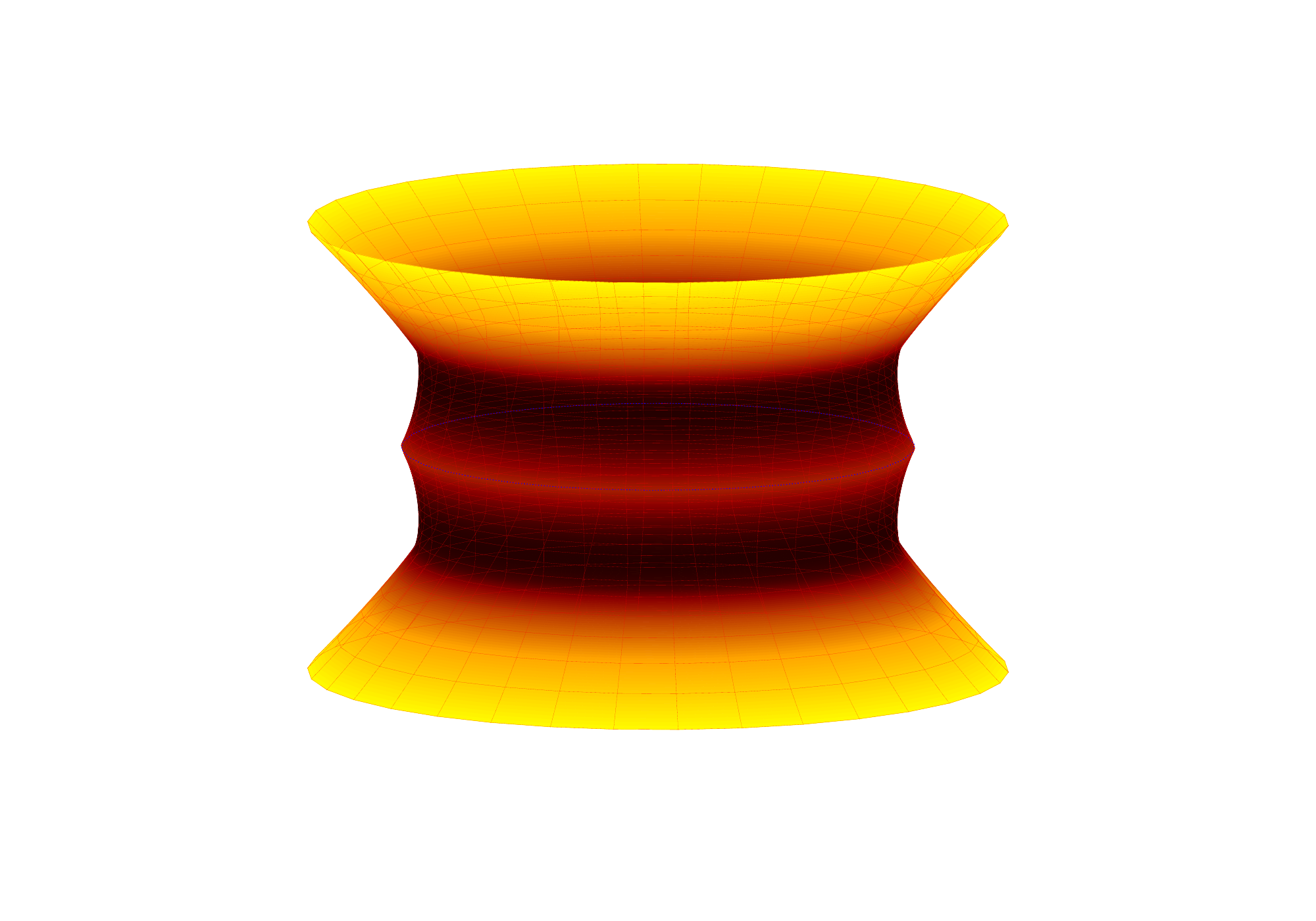}
(b)       \includegraphics[width=.46\textwidth, angle =0]{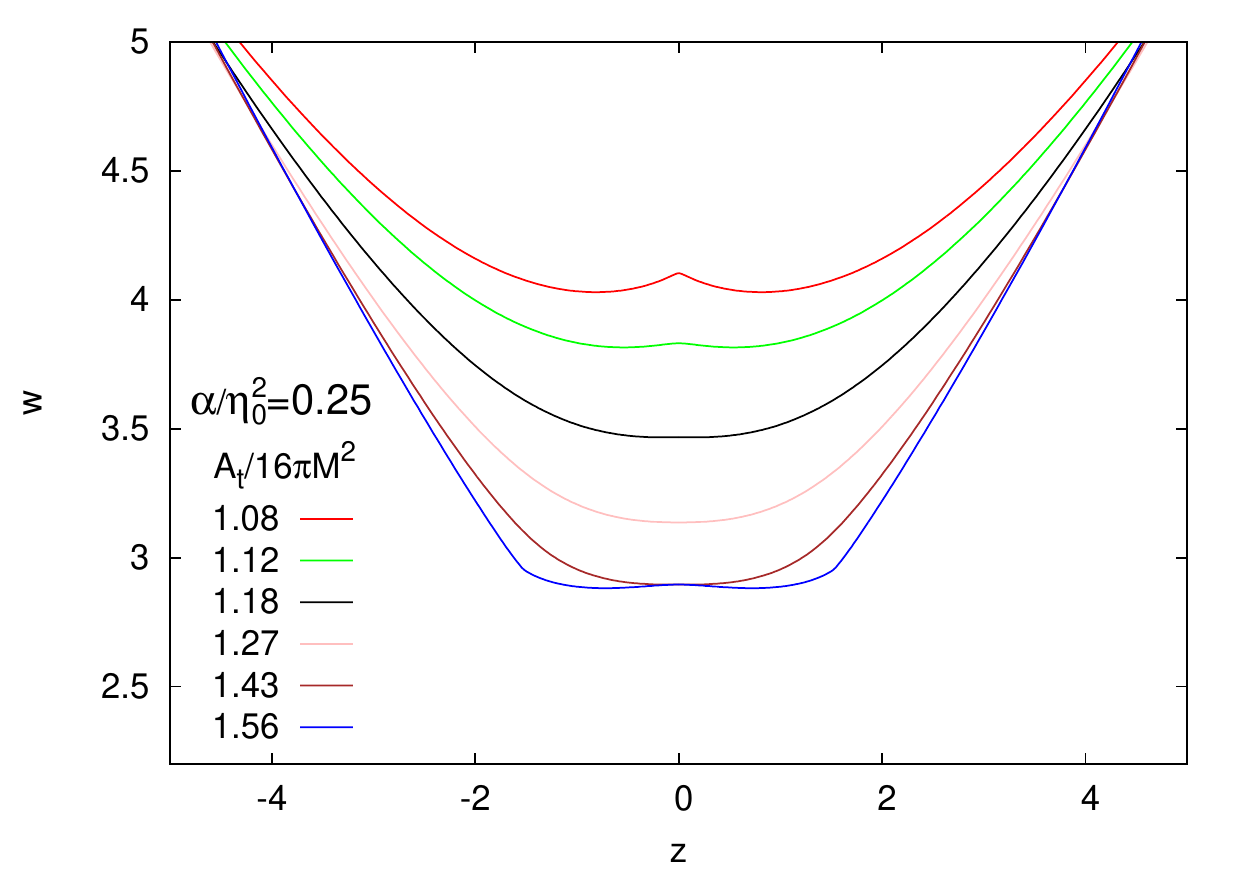}}
\end{center}
\caption{
\label{fig7}
(a) The embedded equatorial plane is shown for the double-throat wormhole with 
$\alpha/\eta_0^2=0.35$ and $D/M=0.886$.
(b) The profiles of the isometric embedding
are shown for a sequence of solutions for the coupling function $F=0.25 e^{-\phi}$.}
\end{figure}

In Fig. \ref{fig7}(a), we depict the isometric embedding of the geometry of a
symmetric, traversable, double-throat wormhole solution. 
The three-dimensional view of the surface follows from the parametric plot
$(w(\eta)\cos\varphi,w(\eta)\sin\varphi,z(\eta))$ as described above.  
The 
diagram clearly features an equator and two throats smoothly connected to two
asymptotic regimes. In Fig.~\ref{fig7}(b), we also show the geometry transition
between single and double-throat wormholes, by plotting $w$ vs. $z$, for a sequence
of solutions for fixed $\alpha/\eta_0^2=0.25$. We observe that, with increasing scaled
throat area, the double-throat wormholes develop a degenerate throat and turn into
single-throat ones. If the scaled throat area is increased further, a second transition
takes place where the single-throat wormholes turn again to double-throat ones.


\section{Conclusions and Discussion}

In this work, we have considered a general class of EsGB theories with an arbitrary
coupling function between the scalar field and the quadratic Gauss-Bonnet term.
By employing a novel coordinate system, we have allowed for wormhole solutions 
with either single-throat or double-throat geometries to emerge. We have determined 
the asymptotic form of the metric functions and scalar field in the small and
large radial-coordinate regimes, and demonstrated that the Null and Weak Energy
Conditions may be violated, especially in the inner regime where the effect of
the GB term is dominant. 

We have then numerically integrated our set of field equations in order to determine
the complete wormhole solutions that interpolate between the derived asymptotic
solutions. We have found wormholes, with either a single throat or a double
throat and an equator, for every form of the coupling function we have tried. 
The spacetime is regular over the entire positive range of the radial coordinate,
as also is the non-trivial scalar field that characterizes every wormhole solution.
Our solutions are therefore characterized by two independent parameters, their mass
and scalar charge. The domain of existence has been studied in detail in each case,
and here we have presented the ones for the exponential and quadratic coupling functions
in order to discuss the qualitative differences as the form of the coupling function
and the value of the scalar field at asymptotic infinity varies. 

An important result of our analysis is that the EsGB theories always feature wormhole
solutions without the need for exotic matter, since the higher-curvature terms
allow for gravitational {\sl effective} negative energy densities. This has been
demonstrated by examining the Null and Weak Energy Conditions for our solutions and
showing that indeed the coupling between the scalar field and the GB term results in
a negative energy density near the throat/equator. The Null Energy Condition is
also violated since it is associated with the appearance of a throat that every
wormhole solution must possess. 

In order to construct traversable wormhole solutions with no spacetime singularities
beyond the throat or equator, our regular solution over the positive range of the
radial coordinate was extended in the negative range in a symmetric way. This
construction demands the introduction of a distribution of matter around the throat
or equator that nevertheless may be shown to consist of physically-acceptable particles.
We have provided an indicative example where this distribution of matter is described
by the equation of state of dust with a vanishing isotropic pressure and a positive
energy density. 

Let us address at this point the issue of the existing bounds on the GB coupling
constant. The parameters of any modified gravitational theory, including the EsGB theory,
may be constrained by processes and observations in strong gravitational regimes. The
most recent bound on the GB coupling parameter $\alpha$ was set in \cite{Nair:2019iur} where
the effect of the scalar dipole radiation on the phase evolution of the gravitational
waveform was taken into account -- this radiation was emitted during the merging process
of two binary systems in which one of the constituents is a scalarised black hole
(GW151226 and GW170608 as detected by LIGO). This bound was set on the value
$\sqrt{\alpha} < 10.1\,{\rm Km}$, taking into account the different definitions of
$\alpha$; in dimensionless units, this translates to $\alpha/M^2<1.72$, where $M$
is the characteristic mass scale of the system, i.e the black-hole mass. In the 
absence of a direct bound on wormholes, since no such object has been detected so far,
and demanding that the EsGB theory should allow for both black-hole solutions and
wormholes to emerge, we apply the aforementioned bound by LIGO on our wormhole
solutions, too. For an exponential coupling function, all of our solutions satisfy
the bound $\alpha/M^2 < 0.91$ while for a quadratic coupling function we obtain
$\alpha/M^2 < 0.605$ (for solutions with no nodes for the scalar field), respectively
$\alpha/M^2 < 2.9$ (for solutions with one node). Thus the observational bound
leaves unaffected the aforementioned DOEs: all solutions in Fig. \ref{fig2}(b) and
Figs. \ref{fig3}(a,b) (with no nodes) fall entirely within the allowed range.

Our next step will be to study the physical characteristics of our solutions in greater
detail and to generalise them to admit also rotation \cite{Kleihaus:2014dla}. In addition,
a linear stability analysis of these EsGB wormholes \cite{Kanti:2011jz,Kanti:2011yv, Evseev:2017jek, Cuyubamba:2018jdl} 
will be performed and their radial and quasi-normal modes, which could be observable
signatures of their existence, will be determined.

\noindent{\textbf{~~~Acknowledgement.--~}}
G.A. would like to thank Onassis Foundation for the financial
support provided through its scholarship program. This research is
co-financed by Greece and the European Union (European Social Fund- ESF)
through the Operational Programme ``Human Resources Development, Education and
Lifelong Learning'' in the context of the project ``Strengthening Human
Resources Research Potential via Doctorate Research'' (MIS-5000432),
implemented by the State Scholarships Foundation (IKY).
BK and JK gratefully acknowledge support by the
DFG Research Training Group 1620 {\sl Models of Gravity}
and the COST Action CA16104. BK acknowledges helpful 
discussions with Eugen Radu. 


\appendix

\section{Field Equations} \label{App-equations}

Employing the metric (\ref{metric}) in Eqs. ~(\ref{einfldeq}), the $(tt)$, $(\eta\eta)$
and $(\theta\theta)$ components of Einstein's equations take the form
\bea
 &\eta _0^6 \left(2 f'_1 \left(\phi ' \left(\dot{F} \left(f'_1{}^2-4
   f''_1\right)-2 f'_1 \ddot{F} \phi '\right)-2 \dot{F} f'_1 \phi ''\right)+e^{f_1}
   \left(f'_1{}^2+4 f''_1+\phi '^2\right)\right)+\eta^4\Big[  
   -4 \dot{F} \eta  f'_1 \phi '' \left(\eta  f'_1+4\right)    \nonumber\\[1mm]
   &+e^{f_1} \eta  \left(\eta  \left(4 f_1''+\phi '^2\right)+\eta 
  f'_1{}^2+8 f'_1\right) + 2 \phi ' \left(\dot{F} \left(f'_1 \left(\eta ^2  f'_1 {}^2-8\right)-4
   \eta  f''_1 \left(\eta  f'_1+2\right)\right)-2 \eta  f'_1 \ddot{F} \phi '
   \left(\eta  f'_1+4\right)\right)\Big]\nonumber\\[1mm]
&+\eta_0^4 \Big[ 2 \phi ' \left(\dot{F} \left(3 f'_1 \left(\eta ^2  f'_1 {}^2-4\right)-4
   \eta  f''_1 \left(3 \eta  f'_1+2\right)\right)-2 \ddot{F} \phi ' \left(\eta 
   f'_1+2\right) \left(3 \eta  f'_1-2\right)\right)-4 \dot{F} \phi '' \left(\eta 
   f'_1+2\right) \left(3 \eta  f'_1-2\right)\nonumber\\[1mm]
&+e^{f_1} \left(\eta  \left(3 \eta  \left(4 f''_1+ \phi '^2\right)+3 \eta
     f'_1 {}^2+8 f'_1\right)+4\right) \Big] +\eta \eta_0^2\Big[-4 \eta  \ddot{F}  \phi '^2 \left(\eta  f'_1 \left(3 \eta 
   f'_1+8\right)-4\right)+e^{f_1}\eta(4 +\eta(16 f_1' \nonumber\\[1mm]
 &+3\eta f_1'^2 +3\eta(\phi'^2+4f_1''))) +   2 \dot{F} \left(\phi ' \left(3 \eta ^3 \left(f'_1\right){}^3-4 \eta ^2 f''_1 
   \left(3 \eta  f_1'+4\right)-20 \eta  f_1'-16\right)-2 \eta  \phi '' \left(\eta 
   f_1' \left(3 \eta  f_1'+8\right)-4\right)\right)=0,\nonumber\\ \label{tteq}
\eea
\bea
  &e^{f_1}\Big[\eta ^3 \left(2 f_0' \left(\eta  f_1'+2\right)+f_1' \left(\eta  f_1'+4\right)-\eta 
   \phi'^2\right)+2 \eta _0^2 \left(\eta  \left(2 f_0' \left(\eta 
   f_1'+1\right)+f_1' \left(\eta  f_1'+2\right)-\eta  \phi'^2\right)-2\right)\nonumber\\
  &+ \eta _0^4 \left(f_1'{}^2+2 f_0' f_1'-\phi'^2\right)\Big]-2 \dot{F} f_0' \phi ' \left(12 \left(\eta ^2+\eta _0^2\right) \eta  f_1'+3
   \left(\eta ^2+\eta _0^2\right){}^2 f_1'{}^2+8 \eta ^2-4 \eta
   _0^2\right)=0,\label{hheq}
\eea
\bea
   &\eta^3\Big[ -4 \dot{F} f_0' \phi '' \left(\eta  f_1'+2\right)-4 f_0' \ddot{F}  \phi'^2 
   \left(\eta  f_1'+2\right)-2 \dot{F} \phi ' \left(-2 f_0' \left(\eta 
    f_1'{}^2-\eta  f_1''+f_1'\right)+2 f_0'' \left(\eta 
   f_1'+2\right)+f_0'{}^2 \left(\eta  f_1'+2\right)\right)\nonumber\\[1mm]
  &+ e^{f_1} \left(\eta  \left(2 \left(f_0''+f_1''\right)+ \phi'^2\right)+\eta   f_0'{}^2+2 f_0'+2 f_1'\right)\Big]
  +2\eta_0^2\Big[ e^{f_1} \left(\eta  \left(\eta  \left(2 \left(f_0''+f_1''\right)+\phi'^2\right)+\eta  f_0'{}^2+f_0'+f_1'\right)+2\right)\nonumber\\[1mm]
  &-2 \dot{F} \phi ' \left(f_0'
   \left(2-\eta  \left(2 \eta  f_1'{}^2-2 \eta 
   f_1''+f_1'\right)\right)+2 \eta  f_0'' \left(\eta  f_1'+1\right)+\eta 
   f_0'{}^2 \left(\eta  f_1'+1\right)\right) -4  \eta  f_0'   \left(\eta  f_1'+1\right)\left(\dot{F}  \phi ''+\ddot{F}
   \phi'^2  \right)\Big]\nonumber\\[1mm]
   &+\eta_0^4\Big[-4 f_0' f_1' \ddot{F}  \phi'^2-4 \dot{F} f_0' f_1' \phi ''-2 \dot{F} \phi ' \left(f_1'
   \left( f_0'{}^2-2 f_1' f_0'+2 f_0''\right)+2 f_0'
   f_1''\right)+e^{f_1} \left( f_0'{}^2+2
   \left(f_0''+f_1''\right)+ \phi'^2\right)\Big]=0, \nonumber\\ \label{ffeq}
\eea
respectively. The scalar equation in turn yields 
\bea
  & \eta _0^6 \left(f_0' \left(4 \dot{F} f_1' f_1''-\dot{F}
   f_1'{}^3+e^{f_1} \phi '\right)+2 \dot{F} f_1'{}^2
   f_0''+\dot{F} f_0'{}^2 f_1'{}^2+2 e^{f_1} \phi
   ''+e^{f_1} f_1' \phi '\right) +\eta^4\Big[ \dot{F} \eta  f_0'{}^2 f_1' \left(\eta  f_1'+4\right)\nonumber\\[1mm]
   &+ f_0' \left(4 \dot{F} f_1' \left(\eta ^2 f_1''+2\right)+\eta  \left( \dot{F} \left(8f_1''  - \eta
    f_1'{}^3   \right)+e^{f_1} \eta  \phi
   '\right)\right)+\eta  \Big(2 \left(\dot{F} \left(\eta  f_1'{}^2 f_0''+4
    f_1' f_0''\right)+e^{f_1} \eta  \phi ''\right)\nonumber\\[1mm]
 &+e^{f_1} \phi ' \left(\eta  f_1'+4\right)\Big] +
 \eta_0^4\Big[ f_0' \left(12 \dot{F} f_1' \left(\eta ^2 f_1''+1\right)-3 \dot{F} \eta ^2
   f_1'{}^3+\eta  \left(8 \dot{F} f_1''+3 e^{f_1} \eta  \phi
   '\right)\right)+6 \dot{F} \eta ^2 f_1'{}^2 f_0''+8 \dot{F} \eta 
   f_1' f_0''\nonumber\\[1mm]
   &-8 \dot{F} f_0''+6 e^{f_1} \eta ^2 \phi '' +\dot{F} f_0'{}^2 \left(3 \eta ^2  f_1'{}^2+4 \eta 
   f_1'-4\right)+e^{f_1} \eta  \phi ' \left(3 \eta  f_1'+4\right)\Big] +\eta\eta_0^2\Big[\dot{F}
   \eta  f_0'{}^2 \left(3 \eta ^2 f_1'{}^2+8 \eta 
   f_1'-4\right) \nonumber\\[1mm]
   &+f_0' \Big(\dot{F} \left(-3 \eta ^3 f_1'{}^3+4
   \eta  f_1' \left(3 \eta ^2 f_1''+5\right)+16 \left(\eta ^2
   f_1''+1\right)\right)+3 e^{f_1} \eta ^3 \phi '\Big)+\eta  \Big(2 \dot{F} f_0'' \left(3 \eta ^2 f_1'{}^2+8 \eta 
   f_1'-4\right)\nonumber\\[1mm]
   &+6 e^{f_1} \eta ^2 \phi ''+e^{f_1} \eta  \phi ' \left(3 \eta 
   f_1'+8\right)\Big)\Big]=0.\label{sceq}
\eea
In the above equations, the prime denotes differentiation with respect to the radial coordinate $\eta$.

\end{document}